\relax
\documentstyle [art12]{article}
\begin {document}
\bibliographystyle {plain}

\title{\bf Anisotropic Spin-1/2 Heisenberg Chain with Open Boundary
Conditions.}
\author {P. de Sa and A. M. Tsvelik}
\maketitle
\begin {verse}
$Department~ of~ Physics,~ University~ of~ Oxford,~ 1~ Keble~ Road,$
\\
$Oxford,~OX1~ 3NP,~ UK$\\
\end{verse}
\begin{abstract}
\par
 We study the exact solution of the anisotropic spin-1/2 Heisenberg
chain with a boundary magnetic field in the region where the bulk
excitations are gapless. It is shown that near the boundary a bound state
is created which is
underscreened when
$\gamma < \pi/3$ and which  at low temperatures behaves  like  a
single spin
weakly coupled to the bulk.  The
IR fixed point in this case belongs to the universality
class of the underscreened anisotropic Kondo model. We argue that the
same fixed point appears in the boundary sine-Gordon model when
the scaling dimension of the boundary term $2/3 < \Delta
< 1$.
\end{abstract}

\sloppy
\par

 Problems about the influence  of impurities and imperfections
on the behaviour of one
dimensional strongly correlated systems are attracting growing
attention. So far the main interest has been
concentrated on effects of potential scattering on the behaviour of
electrons in
Luttinger liquids$^{1,2}$. It is
widely believed that there the conductance of a one-dimensional
chain vanishes at $T = 0$ even if only a single impurity is present,
provided the electron-electron interaction is repulsive $^1$. The best
studied model deals with
spinless interacting electrons and in the continuous limit
is equivalent  to the so-called boundary
sine-Gordon (BSG) model. There are strong  reasons to believe that this
model is related to the problem of the spin-1/2 Heisenberg chain with a
boundary magnetic field$^{3}$:
\begin{equation}
H = \frac{J}{2}\sum_{n = 1}^{N - 1}\left[\frac{1}{2}(- \sigma_n^+\sigma_{n +
1}^-
- \sigma_{n + 1}^+\sigma_{n}^-) + \cos\gamma\sigma_n^z\sigma_{n +
1}^z\right] - \frac{1}{2}h_1\sigma^z_1 - \frac{1}{2}h_2\sigma^z_N
\end{equation}
We shall discuss this relation below. The model (1), however, is
interesting in its own right. At $h_{1,2} = 0$ it describes
an experimentally accessible situation of a one-dimensional magnet
where some spins are replaced by non-magnetic ions.

The  model (1) is exactly solvable by the Bethe ansatz$^{4,5}$.
The Bethe ansatz equations are
\begin{eqnarray}
[e_1(u_a)]^{2N}e_{2S_1}(u_a)e_{2S_2}(u_a) = \prod_{b = 1, b \neq
a}^Me_2(u_a - u_b)e_2(u_a + u_b)\\
e_n(u) = \frac{\sinh[\gamma(u - \mbox{i}n/2)]}{\sinh[\gamma(u +
\mbox{i}n/2)]}\\
\exp(2\mbox{i}\gamma S_j) =
e_1\left[\frac{1}{2\gamma}\ln\left(\frac{h_j}{J} +
\cos\gamma\right)\right] \label{S}
\end{eqnarray}
and the energy is given by
\begin{equation}
E = \bar J\sum_{a =
1}^M\frac{1}{2\mbox{i}\pi}\frac{\mbox{d}}{\mbox{d}u_a}\ln e_1(u_a) -
\frac{1}{2}(h_1 + h_2)
\end{equation}
where
\[
\bar J = \frac{2\pi\sin\gamma}{\gamma}J
\]
The quantities $S_{1,2}$ are defined in such a way that at $h = 0$ $2S
= \pi/\gamma - 1$ and $S = 2\pi/\gamma - 1$ at $h \rightarrow
\infty$. Solutions $u_a$ and $- u_a$ describe the same eigenstate.

 Since we are interested in boundary effects we have to compare the
free energy of the open chain with that of a chain with
periodic boundary conditions. In order to make this comparison we
shall rewrite the Bethe ansatz equations in a form where they are
maximally similar to the equations for the XXZ model with the periodic
boundary conditions. Following Ref. 3, we define a new set of
rapidities $v_a$ such that
\[
v_a = \left\{
\begin{array}{lr}
u_a, & a = 1,2, ... M\\
- u_{2M - a + 1}, & a = M + 1, ... 2M
\end{array}
\right .
\]
Then Eqs.(2) become
\begin{eqnarray}
[e_1(v_a)]^{2N}e_{2S_1}(v_a)e_{2S_2}(v_a)\frac{\sinh[2\gamma(v_a -
\mbox{i}/2)]}{\sinh[2\gamma(v_a + \mbox{i}/2)]} = \prod_{b =
1}^{2M}e_2(v_a - v_b)
\end{eqnarray}
The last term on the left hand side is introduced to compensate the
term with $v_b = - v_a$ now present on the right hand side. This term
can be rewritten as
\begin{equation}
\frac{\sinh[2\gamma(v_a -
\mbox{i}/2)]}{\sinh[2\gamma(v_a + \mbox{i}/2)]} = e_1(v_a)e_{1 -
\pi/\gamma}(v_a)
\end{equation}
The term $e_1(v)$ can be included into the bulk part. Finally, we get
the following Bethe ansatz equations:
\begin{eqnarray}
[e_1(v_a)]^{2N + 1}[e_{2S_1}(v_a)e_{2S_2}(v_a)e_{1 - \pi/\gamma}(v_a)] =
\prod_{b =
1}^{2M}e_2(v_a - v_b) \label{bet}
\end{eqnarray}
with the energy equal to
\begin{equation}
E = - \frac{1}{2}\bar J\sum_{a =
1}^{2M}\frac{1}{2\mbox{i}\pi}\frac{\mbox{d}}{\mbox{d}v_a}\ln e_1(v_a) -
\frac{1}{2}(h_1 + h_2)
\end{equation}

 In these notations the Bethe ansatz equations look similar to  the equations
for the periodic XXZ spin-1/2 Heisenberg chain with an inserted
impurity spin$^6$ or the equations for an anisotropic  Kondo model
with the large coupling constant (see, for example, Ref. 7).
The analogy becomes complete when one of the
boundaries is free $(h_2 = 0, 2S_2 = \pi/\gamma - 1)$. Then the two
phases on the left hand side of Eq.(\ref{bet}) cancel and the
resulting Bethe ansatz equation  is the same as that
for the Kondo model with impurity spin $S_1$. Except for the special case $h_1
\rightarrow \infty$ this spin is never equal to 1/2. Since the XXZ
Heisenberg chain with an impurity spin is very similar to the Kondo
problem$^6$, where the singlet ground state is guaranteed to exist
only if the impurity spin is equal to 1/2, one can expect here
non-analytic contributions from the boundary. As we shall see later,
this  does occur when $0 < \gamma < \pi/2$ when the IR fixed
point of the model (1) coincides with  the fixed point of the
underscreened Kondo model. At $\pi/2 < \gamma < \pi$ the boundary
spins are completely screened at $T \rightarrow 0$.

  Below we shall consider the thermodynamical properties of the model
(1). In order to simplify the calculations  we shall do it at the special
points (a) $\gamma = \pi/\nu$ and (b) $\gamma = \pi(1 - \nu^{-1})$, where
the solutions of Eqs.(\ref{bet}) have especially simple
classification$^8$. The equations for  the distribution functions of
rapidities coincide for the both cases, but the energy has different
signs. Since the results for a finite boundary
magnetic field do not differ qualitatively from those for the open
boundary conditions and the latter case is more physical, we shall
concentrate on the problem with open boundary
conditions and make only few remarks about the general case when it
will be appropriate.

 We emphasise that we are interested in boundary effects, but not in
finite size effects. The latter ones disappear when the length of the chain
becomes infinite. Neglecting finite size effects we get the following
thermodynamic equations for the open boundary conditions:
\begin{eqnarray}
\epsilon_n = - \frac{1}{2}\tilde J\eta s(v)\delta_{n,1} + Ts*\ln\left(1 +
\mbox{e}^{\epsilon_{n - 1}/T}\right)\left(1 +
\mbox{e}^{\epsilon_{n + 1}/T}\right)\nonumber\\
+ T\delta_{n,\nu - 2}s*\ln
\left(1 + \mbox{e}^{\epsilon_{0}/T}\right), \: n = 1, ... \nu - 2\\
\epsilon_{\nu - 1} = Ts*\ln\left(1 + \mbox{e}^{\epsilon_{\nu - 2}/T}\right) +
\frac{\nu}{2}H\\
\epsilon_{0} = Ts*\ln\left(1 + \mbox{e}^{\epsilon_{\nu - 2}/T}\right) -
\frac{\nu}{2}H\\
F_{bulk} = - 2N T\int_0^{\infty}\mbox{d}v s(v)\ln\left(1 +
\mbox{e}^{\epsilon_{1}/T}\right)\\
F_{boundary} = F_1 + F_2\nonumber\\
F_1 = - T\int_0^{\infty} \mbox{d}v s(v)\ln\left(1 +
\mbox{e}^{\epsilon_{\nu - 1}/T}\right)\left(1 +
\mbox{e}^{\epsilon_{0}/T}\right)\nonumber\\
F_2 = - T\int_0^{\infty}\mbox{d}v s(v)\ln\left(1 +
\mbox{e}^{\epsilon_{1}/T}\right)\label{fboun}
\end{eqnarray}
where $\eta = 1$ for $\gamma < \pi/2$ and $\eta = - 1$ for $\gamma >
\pi/2$ and  H is the magnetic field in the bulk.
The sign $*$ stands for convolution
\[
f*g(v) = \int_{-\infty}^{\infty}f(v - v')g(v')\mbox{d}v'
\]
and $s(v) = [4\cosh(\pi v/2)]^{-1}$.

 As we have said, one can expect these equations to be similar to the
equations for the Kondo problem. There is one important difference,
however: the non-trivial part of the boundary
free energy $F_2$ (\ref{fboun}) is equal to 1/2 of the free
energy of the Kondo impurity. This is, of course, due to the
restriction that only symmetric distribution of rapidities  are
allowed. Eqs.(10-14) are also similar to the
thermodynamic equations for BSG problem derived
in Refs. (9, 10) from the bootstrap solution of Ref. 11. The
anisotropy parameter $\gamma$ is related to the scaling dimension of
the boundary term in BSG
problem:
\begin{equation}
\Delta = 1 - \gamma/\pi
\end{equation}
 Here the
differences
are more important. The least  important one  is that  the BSG equations
contain the boundary energy scale  $T_B$. This difference
by setting  $T_B$ equal to the
ultraviolet cut-off $\Lambda \approx J$. What is more important,
however,
is the fact
that the boundary free energy in BSG model has a different form:
\begin{equation}
F_{BSG} =  - T\int_{-\infty}^{\infty} \mbox{d}v s[v -
\frac{2}{\pi}\ln(\Lambda/T_B)]\ln\left(1 +
\mbox{e}^{\epsilon_{\nu - 1}/T}\right)
\end{equation}
At $H = 0$ when $\epsilon_{\nu - 1} = \epsilon_0$ and $\Lambda = T_B$
these two expressions are equivalent, but at $H \neq 0$ one can expect
differences. These differences are due to the assymetry between solitons and
antisolitons in BSG.

 Since the obtained thermodynamic equations are very similar to those
for BSG model and the latter were analysed for $\eta = -1$ in
Refs. (9, 10), we shell concentrate on the case $\eta = + 1, \: \gamma
< \pi/2$. In this case the ground state consists of
real $v$'s, i.e. the only non-vanishing energy at $T = 0$ is
$\epsilon_1$.

 Analytical solutions are available for asymptotics of $\epsilon_n(x)$
at $v \rightarrow 0, \infty$ (see, for example, Refs.(7, 8)).
At large temperatures $T >> J$ the free energy is determined by the
asymptotics at $v \rightarrow + \infty$ where
\begin{eqnarray}
\left(1 + \mbox{e}^{\epsilon_{0}/T}\right)\left(1 +
\mbox{e}^{\epsilon_{\nu - 1}/T}\right) = \left[\frac{\sinh \nu H/2T}{\sinh
H/2T}\right]^2
\end{eqnarray}
such that we have
\begin{equation}
F_{boundary} \rightarrow - \frac{T}{2}\ln\left[\frac{\sinh \nu H/2T}{\sinh
H/2T}\right]
\end{equation}
At small temperatures the leading contribution comes from the region $v
\rightarrow  0$ where $\epsilon_n \: (\nu \neq 1)$ are again almost
constant and $\exp(\epsilon_1/T)$ is small. Then the
corrections can be determined from the expansion in  $\exp(\epsilon_1/T)$:
\begin{eqnarray}
g_n(v) \equiv \ln\left(1 + \mbox{e}^{\epsilon_{n}(v)/T}\right) =
g_n^{(0)} +
g^{(1)}_n(v) + ... \nonumber\\
g_n^{(0)} = 2\ln\Phi(n) \: (n = 2,3, ... \nu - 2), \nonumber\\
g_{\nu - 1}^{(0)} =
\ln[1 + \exp(\nu H/2T)\Phi(\nu - 2)], \:  g_{0}^{(0)} =
\ln[1 + \exp(- \nu H/2T)\Phi(\nu - 2)]\nonumber\\
g^{(1)}_n(v) = \frac{1}{\Phi(2)\Phi(n)}[\Phi(n + 1) a_n*g_1(v) - \Phi(n -
1)a_{n + 2}*g_1(v)]\\
\Phi(n) = \frac{\sinh n\nu H/2(\nu - 1)T}{\sinh \nu H/2(\nu - 1)T}, \:
a_n(\omega) = \frac{\sinh[(\nu - n)\omega/2]}{\sinh[(\nu - 1)\omega/2]}
\end{eqnarray}
Using these expressions we get the following expansion for the free energy:
\begin{eqnarray}
F_{boundary} \rightarrow - \frac{T}{2}\ln\Phi(\nu - 1) -
T\int_{0}^{\infty} \mbox{d}v
f(v)\ln\left(1 + \mbox{e}^{\epsilon_1(v)/T}\right)
\end{eqnarray}
where $\Delta = 1 - 1/\nu = 1 - \gamma/\pi$ and
\[
f(\omega) = \frac{\tanh(\omega/2)}{\sinh[(\nu - 1)\omega/2]}
\]
In order to estimate the second term in Eq.(21) we use the crude
approximation:
\begin{equation}
\ln\left(1 + \mbox{e}^{\epsilon_1(v)/T}\right) \approx C\theta\left(v
- \frac{2}{\pi}\ln(J/T)\right)
\end{equation}
where $C$ is a constant.

The first result is that the ground state has a finite entropy $S(0) =
- \frac{1}{2}\ln(\Delta^{-1} - 1)$ which corresponds to the the half of
the entropy of the
underscreened spin $(S - 1/2)$. A careful analysis shows that this
entropy disappears at $\Delta \leq 1/2$.
The ratio of boundary contributions to the partition functions in
the ultraviolet and the infrared limits is
\begin{equation}
Z_{UV}/Z_{IR} = \Delta^{-1/2}
\end{equation}
The same expression remains valid at $\pi > \gamma > \pi/2$$^9$.

 From the obtained expression for the free energy we derive
the following asymptotics for the
specific heat and the magnetic susceptibility at $\nu > 2, \: H
\rightarrow 0$:
\begin{eqnarray}
\chi = \frac{(\nu +  1)\nu^2}{24(\nu - 1)T}\left[1 +
B\left(\frac{T}{J}\right)^{2/(\nu - 1)} + ... \right]\\
C_v \sim (T/J)^{2/(\nu - 1)} \label{eq:C}
\end{eqnarray}
where $B$ is a constant.

 Now we shall  calculate the boundary
contribution to the overall magnetic moment $M^{boundary}$
and the average value of
spin on the boundary $\langle S^z_1\rangle$ at $T = 0$. At
zero temperature  the
only non-vanishing density is the density of real $v$'s $\rho(v)$. From the
Bethe ansatz equations (8) we derive the integral equation for
$\rho(v)$:
\begin{eqnarray}
\int_{-B}^B \mbox{d}u A_{11}(v - u)\rho(u) = s*A_{11}(v) +
\frac{1}{2N}\mu_1(v)\label{rho}\\
E = - \frac{h_1}{2} - \bar J N\int_{-B}^B \mbox{d}v s*A_{11}(v)\rho(v)
\label{E}
\end{eqnarray}
where
\[
A_{11}(\omega) = 1 + \frac{\sinh(\nu - 2)\omega/2}{\sinh\nu\omega/2}, \:
\mu(\omega) = \frac{\sinh(\nu - 2S)\omega/2}{\sinh\nu\omega/2}
\]
and the limit $B$ is determined by the magnetic field in the bulk such
that at it is infinite at $H = 0$. Let us consider the case $h = 0$
first. At $H << \bar J$ the Fredholm equation (\ref{rho}) can be
treated as a Wiener-Hopf equation. The latter can be solved
giving  the leading asymptotics of the magnetization.  The result is
\begin{eqnarray}
M^{boundary}  =
\frac{1}{4\pi}\int_{-\infty}^{\infty}\mbox{d}x \frac{\exp[-
\frac{2\mbox{i}x}{\pi}\ln(\bar J/H)]}{x -
\mbox{i}0}\frac{1}{G^{(-)}(x)}\frac{\sinh
x}{\sinh[\Delta x/(1 - \Delta)]\cosh x} \nonumber\\
\approx \frac{1}{2\sqrt\pi}\tan(\pi/\Delta)\frac{\Gamma(1 +
1/\Delta)}{\Gamma(1/\Delta - 1/2)} (H/\bar J)^{2(\Delta^{-1} - 1)}
\nonumber\\
+
\frac{1}{\sqrt\pi\cos[\pi/2(1 - \Delta)]}\frac{\Gamma[1 + 1/2(1 -
\Delta)]}{\Gamma[1 + \Delta/2(1 - \Delta)]}(H/\bar J)
\label{eq:M}
\end{eqnarray}
where
$G^{(\pm)}(x)$ are  functions analytical in the upper (lower)
half-plane of $x$:
\begin{equation}
G^{(+)}(-x) = G^{(-)}(x) =  \frac{\Gamma(1 +
\frac{\mbox{i}x\Delta}{(1- \Delta)\pi})\Gamma(\frac{1}{2} +
\frac{\mbox{i}x}{\pi})}{\Gamma(1 + \frac{\mbox{i}x}{\pi(1
- \Delta)})\Gamma(1/2)}
\end{equation}
At $\Delta \geq 2/3 \: (\nu \geq 3)$ the non-analytic contribution
becomes dominant. At $\Delta = 2/3$ there is a double pole and
we have a marginal situation with $M^{boundary} \sim H\ln H$. Comparing
Eqs. (\ref{eq:M}) and (\ref{eq:C}) we see that H has the same
scaling dimension as T which is quite natural since if we have an
unscreened spin in our theory.

 It is also interesting to know whether the underscreened spins are
situated directly on  the
boundaries. To find this out  we calculate the average
value of the boundary
spin explicitly at $H = 0, \: T = 0$. From Eqs.(\ref{rho}) and
(\ref{E}) we find the $h_1$-dependent part of the ground state energy:
\begin{equation}
E(h) = - \frac{h_1}{2} -
J\frac{\nu\sin(\pi/\nu)}{\pi}\int_0^{\infty}\mbox{d}\omega\frac{\sinh[\nu
- 2S(h_1)]\omega}{\sinh\nu\omega\cosh\omega}
\end{equation}
Differentiating this expression in $h_1$ and using the definition of $S$
(\ref{S}) we get
\begin{equation}
 \langle S^z_1\rangle = - \frac{\partial E}{\partial h_1}|_{h_1 =
0} = \frac{1}{2} -
\frac{2}{\pi^2}\int_0^{\infty}\mbox{d}\omega\frac{\omega}{\sinh\omega}
= 0
\end{equation}
and
\begin{equation}
 \chi_{boundary} = - \frac{\partial^2 E}{\partial h_1^2}|_{h_1 =
0} = \frac{4}{\gamma^3\sin\gamma J}\int_0^{\infty}\frac{\mbox{d}\omega
\omega^2\tanh\omega}{\sinh(\pi\omega/\gamma)}
\end{equation}
Thus we see that  the susceptibility of the boundary spins is finite. This
means that the boundary spins do not participate in the bound state
responsible for the singularities in the free energy.

 In conclusion, as we have suggested above, at $\gamma < \pi/3$
the boundary contribution to the
free energy is non-analytic in T and H and the boundary free energy is
equal to 1/2 of the free energy of the corresponding underscreened
Kondo model. Since the XXZ chain is similar to BSG model, we expect
the same type of behaviour for BSG in the area $1/2 < \Delta < 1$.

The authors gratefully acknowledge important conversations with
P. Coleman, M. Evans, L. Ioffe, E. Fradkin and I. Kogan.

\end{document}